\documentclass[preprint,amsmath,amssymb,superscriptaddress]{revtex4}

\usepackage{chemformula} 
\usepackage[T1]{fontenc} 
\usepackage{amsmath}
\usepackage{amsfonts}
\usepackage[labelfont=bf]{caption}
\usepackage{color}
\usepackage{float}
\usepackage{atbegshi}
\usepackage{upgreek}
\usepackage{xcolor}
\usepackage{multirow}
\usepackage{placeins}
\usepackage[normalem]{ulem}
\usepackage{graphicx}
\usepackage{xr}
\usepackage{braket}

\newcommand{\x}[1]   {\mathrm{#1}}

\begin{document}

\title{Dielectric control of ultrafast carrier dynamics and transport in graphene}

\author{{Hai I.} {Wang}} 
\email{k.j.tielrooij@tue.nl, h.wang5@uu.nl}
\affiliation{Max Planck Institute for Polymer Research, Ackermannweg 10, Mainz, 55128, Germany}
\affiliation{Debye Institute for Nanomaterials Science, Utrecht University, Princetonplein 1, Utrecht, 3584, The Netherlands}

\author{{Xiaoyu} {Jia}}
\affiliation{Max Planck Institute for Polymer Research, Ackermannweg 10, Mainz, 55128, Germany}

\author{{Anand} {Nivedan}}
\affiliation{Catalan Institute of Nanoscience and Nanotechnology (ICN2), CSIC and BIST, Campus UAB, Bellaterra, 08193, Barcelona, Spain}

\author{{Mischa} {Bonn}}
\affiliation{Max Planck Institute for Polymer Research, Ackermannweg 10, Mainz, 55128, Germany}

\author{{Aron W.} {Cummings}}
\affiliation{Catalan Institute of Nanoscience and Nanotechnology (ICN2), CSIC and BIST, Campus UAB, Bellaterra, 08193, Barcelona, Spain}

\author{{Alessandro} {Principi}}
\affiliation{School of Physics and Astronomy, University of Manchester, UK}

\author{{Klaas-Jan} {Tielrooij}}
\email{k.j.tielrooij@tue.nl, h.wang5@uu.nl}
\affiliation{Department of Applied Physics, TU Eindhoven, Den Dolech 2, Eindhoven, 5612 AZ, Netherlands}
\affiliation{Catalan Institute of Nanoscience and Nanotechnology (ICN2), CSIC and BIST, Campus UAB, Bellaterra, 08193, Barcelona, Spain}

\begin{abstract} 
\textbf{Understanding the ultrafast dynamics of photoexcited charges in graphene is essential, as the microscopic mechanisms underlying these dynamics determine many of graphene’s optical, optothermal, and optoelectronic properties. These are crucial properties for many functionalities and devices enabled by graphene, such as high-speed photodectors. Therefore, beyond scientific understanding, it is highly desirable to control ultrafast carrier dynamics for practical applications. Here, we establish this control by engineering the dielectric environment of graphene, thereby regulating both heating and cooling dynamics without altering the Fermi energy, optical power, or ambient temperature. By combining optical pump – terahertz probe experiments with theoretical calculations, we show that dielectric screening suppresses carrier-carrier interactions and slows the dynamics. In particular, reduced carrier-carrier scattering delays the formation of a quasi-equilibrium hot electron distribution, thus slowing carrier heating. It also slows carrier cooling because re-thermalization after optical-phonon emission depends on the same interactions. The enhanced screening further reduces the energy of electron-hole puddles, thereby increasing charge mobility and the Seebeck coefficient. This ability to externally control internal graphene dynamics and transport properties enables the optimization of device performance, such as the sensitivity of photodetectors for data communication and wireless communication applications.
}
\end{abstract}
\maketitle

\newpage

\section*{Introduction}

The optoelectronic properties of graphene enable many exciting functionalities and devices, such as ultrafast photodetectors and receivers that operate from the visible through the telecom up to the (sub)terahertz regime \cite{Lemme2011, Song2011, Cai2014, Schuler2016, Muench2019, Cummings2021, Koepfli2024, Vicarelli2012b,Castilla2019,Viti2020,Soundarapandian2026}, nonlinear optical converters \cite{Hafez2018,Soavi2018a,Jiang2018}, saturable absorbers for lasers \cite{Martinez2013,Autere2018,Tan2020}, and light emitters \cite{kim2015, luo2019, Shiue2019}. The ultrafast dynamics of photoexcited carriers in graphene have therefore been studied in depth, and are relatively well understood \cite{Massicotte2021}. First, the initially photoexcited carriers undergo thermalization via electron-electron interactions, which leads to a hot electron distribution with an electron temperature that is larger than the lattice temperature. The second step is electron cooling via electron-phonon thermalization. The first step -- electron heating -- takes a few tens of femtoseconds \cite{Brida2013,Gierz2013,Tielrooij2013}, depending on the excitation wavelength, excitation power density, and Fermi energy \cite{Song2013, Jensen2014}. The second step -- electron cooling -- occurs on a picosecond timescale at room temperature. Electron-phonon cooling can involve strongly coupled optical phonons \cite{Rana2009, Breusing2011} optical-to-acoustic phonon cooling \cite{Wang2010, Malard2013}, disorder-assisted coupling to acoustic phonons \cite{Song2012,Graham2013}, or direct coupling between hot electrons and substrate phonons, for example to hyperbolic phonon polaritons in hBN \cite{Yang2018, Tielrooij2018} or to molecular modes in water \cite{Yu2023}. In the absence of a substrate that acts as an efficient heat sink for hot electrons, and in the case of relatively high charge mobility, cooling is ultimately dominated by a continuous process of optical phonon emission and electron re-thermalization that leads to secondary hot electrons with enough energy to emit optical phonons \cite{Pogna2021}. 
\\

Several cooling mechanisms have some degree of tunability, because they depend -- to some extent -- on the Fermi energy, which can be controlled electrically \cite{Graham2013a, Tielrooij2018, Pogna2022}. Moreover, the cooling dynamics can be influenced by the incident optical power due to an optical-to-acoustic phonon bottleneck that leads to reheating of the electron system by phonons \cite{Wang2010,Pogna2021}. Finally, the cooling dynamics depend on the ambient temperature: at cryogenic temperatures, diffusive cooling, where electronic heat diffuses out of the heated region, can become dominant, in particular for micrometer-scale heated areas, because electron-phonon cooling channels become less efficient at reduced temperatures \cite{Crossno2016,Block2021}. For many applications, it would be beneficial to have a way to control both the cooling and heating dynamics without changing the Fermi energy, optical power, and ambient temperature. It would be particularly useful to increase the hot-carrier cooling time, as this would lead to a larger photoresponse and therefore a higher sensitivity of photo-thermoelectric devices. However, such a mechanism for controlling carrier dynamics is currently unavailable, as the only strategy appears to be modifying the intrinsic optical and acoustic phonon properties of graphene.
\\

 \begin{figure*}[h!]
  \centering
    \includegraphics[width=0.95\linewidth]{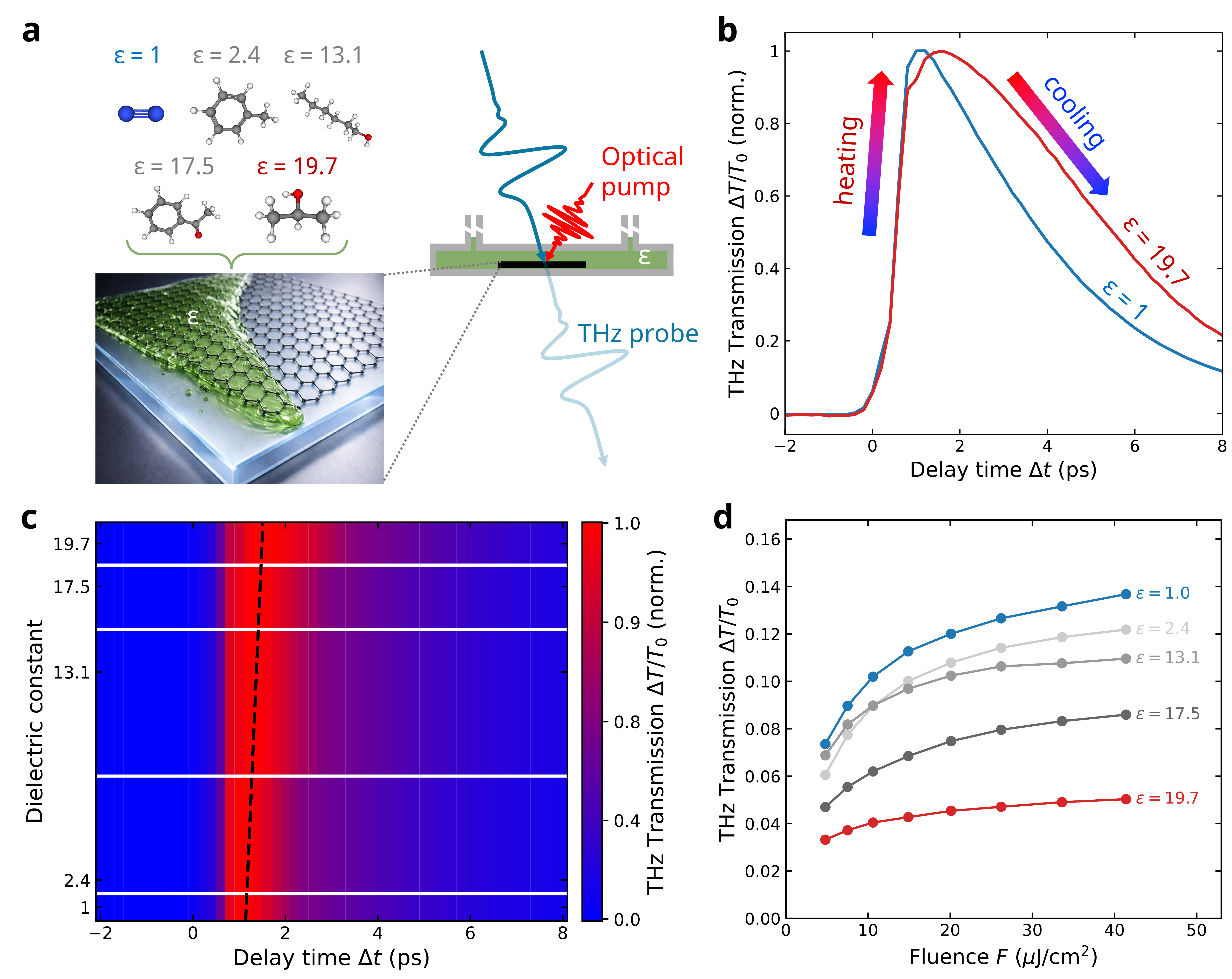}
    \caption{\textbf{Experimental observation of slower photoexcited carrier dynamics controlled by the dielectric environment.} \textbf{a,} Illustration of the experimental approach, where graphene inside a modified silica flow cell is surrounded by different dielectric environments with static dielectric constant $\epsilon$ (in green), in particular nitrogen gas, toluene, 1-hexanol, acetophenone, and isopropanol (IPA), see chemical structures on the top left. Time-resolved optical pump -- THz probe measurements provide access to the ultrafast carrier dynamics, as shown on the right. \textbf{b,} The pump-induced change in THz transmission $\Delta T/T_0$ as a function of pump-probe delay time $\Delta t$ for nitrogen ($\epsilon$ = 1) and IPA ($\epsilon$ = 19.7) environments, showing a markedly slower rise and slower decay for the IPA case. \textbf{c,} The photoexcited carrier dynamics for the different environments, with the dashed line indicating the delay time where the peak signal occurs. The results show a gradual slowing of the photoexcited carrier dynamics -- both heating and cooling -- as the dielectric constant of the environment increases. \textbf{d,} Peak THz transmission as a function of pump fluence for the different dielectric environments, showing a lower signal and faster saturation for larger dielectric constants. }
    \label{fig:exp}
\end{figure*}

Here, we demonstrate the controllability of the photoexcited carrier dynamics of monolayer graphene using dielectric engineering, without modifying any phonon properties. The mechanism we identified acts on the thermalization process for heating dynamics and the re-thermalization process for cooling dynamics. Specifically, by engineering the dielectric environment, we control the screening of electron-electron interactions \cite{Kim2020}, which in turn governs the (re-)thermalization dynamics. Employing optical pump -- terahertz (THz) probe measurements, we experimentally demonstrate slower heating and cooling dynamics due to increased screening, as induced by liquid superstrates with varying dielectric constants (see Fig.~\ref{fig:exp}a). We developed a theoretical model that describes the effect of screening on photoexcitation dynamics and confirms the experimentally observed trends. We also calculated the effect of screening on the electronic and thermoelectric properties of graphene, finding that the charge mobility and Seebeck coefficient increase with stronger screening. Finally, we demonstrate that these effects predict an increase in sensitivity for photo-thermoelectric photodetectors. 
\\

\section*{Results}
\subsection*{Experimental signatures of dielectric control of ultrafast carrier dynamics}

We studied ultrafast photoexcited carrier dynamics of large-area graphene, which was grown by chemical vapour deposition (CVD) and transferred onto a SiO$_2$ substrate. This substrate served as a detachable window for a liquid flow cell, through which we could flow different liquids that served as a dielectrically controllable superstrate for graphene (see Fig.~\ref{fig:exp}a). This enabled us to conduct time-resolved THz measurements on the same sample location while tuning the liquid dielectric environment. The specific liquids that we used were toluene, 1-hexanol, acetophenone, and isopropanol (IPA), with dielectric constants $\epsilon$ of 2.4, 13.1, 17.4, and 19.7, respectively \cite{Madelung1991}. We characterized the graphene sample under different dielectric environments using Raman microscopy (see Supp.~Figs.~S1-S2). Before each measurement, we first measured the dynamics in the absence of solvent under a purged atmosphere of dry nitrogen, corresponding to a dielectric constant of $\sim$1.  To monitor the dynamics of photoexcited carriers, we excited graphene using laser pulses with a pulse duration of $\approx$35 fs and a wavelength of 800 nm (photon energy $\hbar \omega$ = 1.55 eV), and probed the electron system using quasi-single-cycle terahertz (THz) pulses with a photon energy of 1-10 meV. These THz pulses are directly sensitive to the photoexcited charge carriers via their Drude conductivity. Specifically, photo-induced electron heating for graphene with the Fermi energy away from the Dirac point leads to a reduced Drude conductivity, \textit{i.e.}\ negative photoconductivity, and therefore increased THz transmission \cite{Tielrooij2013,Frenzel2014,Wang2017a,Tomadin2018}. Scanning the temporal delay between the optical pump and THz probe pulses using an optical delay line provides access to the ultrafast photoexcited electron dynamics. For more experimental details, see Methods:Experimental details. 
\\

Figure \ref{fig:exp}b shows two exemplary measurements of the pump-induced change in THz transmission as a function of pump-probe delay time $\Delta t$, for the case of a nitrogen environment ($\epsilon$ = 1) and an IPA environment ($\epsilon$ = 19.7). The increase in transmission directly after $\Delta t$ = 0, where the pump and probe pulses overlap, typically corresponds to carrier heating, while the subsequent decay corresponds to carrier cooling. Clearly, both the heating and cooling dynamics are significantly slower for the IPA case. If we systematically increase the dielectric constant using different liquids, we observe a gradual slowing of the heating and cooling dynamics, as shown in Fig.~\ref{fig:exp}c. In addition, the pump-probe signal amplitude decreases for larger dielectric constants, as shown in Fig.~\ref{fig:exp}d (see Supp.~Fig.~S3 for an overview of the experimental data). Our Raman measurements show that the different environments do not lead to drastic changes in carrier density (see Supp.~Figs.~S1-S2). Moreover, slower dynamics would imply a larger Fermi energy \cite{Pogna2022}, which would, in turn, lead to a larger pump-probe signal. In contrast, we observed a reduced pump-probe signal combined with slower dynamics. Therefore, we conclude that the observed changes in the photoexcited carrier dynamics and the pump-probe signal are not due to a shift in the Fermi energy.
\\

\subsection*{Tuning ultrafast carrier dynamics through dielectric screening}

\begin{figure*}[!htp]
  \centering
    \includegraphics[width=\linewidth]{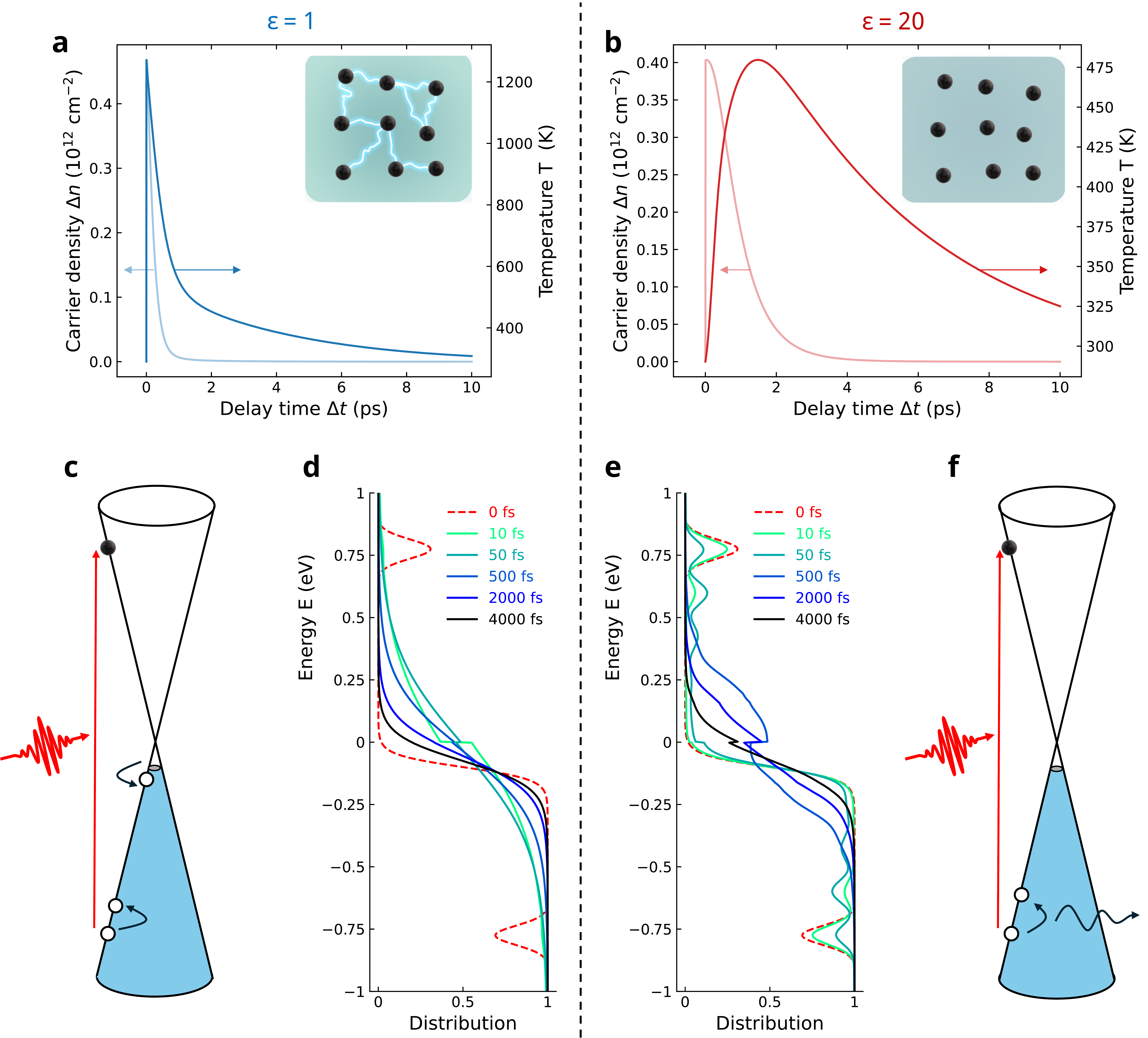}
    \caption{\textbf{Calculated dielectric tuning of photoexcited carrier dynamics by controlling carrier-carrier interactions.} \textbf{a-b,} Calculated photoexcited carrier dynamics in an environment with $\epsilon$ = 1 \textbf{(a)} and $\epsilon$ = 20 \textbf{(b)}, showing the dynamics of the density of photoexcited carriers $\Delta n$ and of the carrier temperature $T$. The latter is obtained by describing the corresponding carrier distributions with Fermi-Dirac statistics. \textbf{c,} Illustration of the carrier dynamics that occur after optical excitation creates an additional electron in the conduction band and an additional hole in the valence band, for $p$-doped graphene with $\epsilon$ = 1. The initially excited carriers relax through carrier-carrier scattering, which creates additional hot carriers, in this case, hot holes with an energy above the Fermi energy. \textbf{d-e,} Snapshots of calculated carrier distributions at several delay times after photoexcitation for $\epsilon$ = 1 \textbf{(d)} and $\epsilon$ = 20 \textbf{(e)}. In the former case, all distributions after $\approx$10 fs are thermalized due to efficient carrier-carrier interactions, whereas in the latter case, they are non-thermal. \textbf{f,} Illustration of the carrier dynamics after photoexcitation for $\epsilon$ = 20. In this case, the relaxation of initially excited carriers by the emission of optical phonons plays an important role, since carrier-carrier interactions are screened.  }
    \label{fig:theory_dynamics}
\end{figure*}

We hypothesize that dielectric environments affect carrier dynamics by screening electron-electron interactions. The idea is that environments with a higher dielectric constant lead to increased screening of electron-electron scattering \cite{Kim2020}, which in turn affects the photoexcited carrier dynamics. We test this hypothesis using an analytical model that considers the cooling dynamics described in Ref.~\cite{Pogna2021}, occurring through a combination of optical phonon emission and carrier re-thermalization via electron-electron scattering. We extended this model to include different dielectric environments (see Supplementary Note 1), focusing on two cases: $\epsilon$ = 1 and $\epsilon$ = 20. The results shown in Fig.~\ref{fig:theory_dynamics}a-b demonstrate that a larger dielectric constant indeed gives rise to a slower rise in electron temperature and a slower decay as well, in agreement with the experimental observations. The simulations predict a more pronounced slowing than what we observed in the experiment. One reason is that in the experiment, only the dielectric environment above the graphene changes, whereas the simulation assumes that the entire dielectric environment changes. Furthermore, since the experiment used CVD-grown graphene, cooling can occur via an additional channel: disorder-assisted acoustic phonon emission \cite{Song2012}, which sets an upper limit on the cooling time. 
\\

In order to understand the pump-induced carrier dynamics in different dielectric environments in more detail, we examine the processes that occur after the absorption of pump light creates initially excited charge carriers, which are additional electrons in the conduction band and additional holes in the valence band at energies $\pm \frac{1}{2} \hbar \omega$, see Fig.~\ref{fig:theory_dynamics}c. This situation corresponds to a non-thermal distribution at $\Delta t$ = 0. In this simulation, we assume that the Fermi energy is located below the Dirac point, meaning that holes are the majority charge carriers. The behaviour would be identical for a Fermi energy above the Dirac point with electrons as majority carriers. Figure \ref{fig:theory_dynamics}d shows the evolution of the carrier distribution as a function of time after photoexcitation for the case $\epsilon$ = 1. Within several tens of femtoseconds, a thermalized Fermi-Dirac distribution is established, with an elevated carrier temperature. This occurs through a cascade of carrier-carrier interactions starting from the initially photoexcited carriers, which leads to the creation of additional hot carriers (see Fig.~\ref{fig:theory_dynamics}), \cite{Tielrooij2013, Wu2016f}. Within several picoseconds, the system relaxes back to equilibrium with an electron temperature equal to the lattice temperature. By describing the carrier distribution with a Fermi-Dirac distribution characterized by a temperature and a chemical potential, we obtain the time-dependent carrier temperature $T$ shown in Fig.~\ref{fig:theory_dynamics}a. We also obtain the net photo-induced carrier density $\Delta n$ as a function of time, and observe that within a picosecond, there are no net photo-induced charge carriers left in the system, which means that the system is purely characterized by an increased carrier temperature. The photoexcitation dynamics for graphene in an environment with $\epsilon$ = 1 are therefore similar to the dynamics of a typical metal, giving a negative photoconductivity and dynamics that reflect the heating-cooling dynamics of the electronic system. 
\\

The evolution of the carrier distribution is very different for $\epsilon$ = 20. In this case, the carrier distributions are much more non-thermal, as shown in Fig.~\ref{fig:theory_dynamics}e. Rather than a continuous cascade of carrier-carrier interactions leading to a thermal distribution, there is a significant contribution of optical phonon emission to the relaxation of the initially photoexcited carriers, as illustrated in Fig.~\ref{fig:theory_dynamics}f. This leads to slower and less efficient carrier heating. Using a Fermi-Dirac distribution with an ``effective'' carrier temperature, we obtain the carrier temperature dynamics in Fig.~\ref{fig:theory_dynamics}b. We observe slower dynamics and a lower temperature for $\epsilon$ = 20 than for the case of $\epsilon$ = 1. The lower temperature explains the experimentally observed decrease in pump-probe signal, which scales with the carrier temperature. We also find that the net photo-induced carrier density $\Delta n$ lives significantly longer for $\epsilon$ = 20. This likely explains the smaller signal magnitude observed in Fig.~\ref{fig:exp}d, because an increased carrier density gives rise to a positive photoconductivity, \textit{i.e.} a decrease in transient transmission. The photoexcitation dynamics for graphene in an environment with $\epsilon$ = 20 are therefore more similar to the dynamics of a typical semiconductor. Importantly, the combination of these simulations and our experimental results demonstrate the ability to externally control the cooling dynamics driven by intrinsic electron-phonon interactions in graphene. 
\\

\subsection*{Dielectric tuning of transport properties}

Next, we use numerical simulations to examine how altering the dielectric environment can tune the electronic transport properties of graphene. To quantify this, we assume that in graphene on SiO$_2$, charge transport is limited by electron-hole puddles. These arise from charges trapped in the oxide, inducing a spatially varying electrostatic potential in the graphene layer that can dominate charge transport \cite{Adam2009, Klos2010}. Scanning tunneling spectroscopy (STS) measurements have revealed that this potential has a Gaussian distribution with a standard deviation of $50$ meV and a length scale of $\sim$$10$ nm \cite{Deshpande2009, Xue2011}.
\\

To simulate charge transport in graphene with disorder on such a length scale, we employ a linear-scaling real-space quantum transport method capable of handling systems with many millions of atoms (see Methods for details) \cite{Fan2021}. With this we calculate the Fermi-energy-dependent electrical conductivity, $\sigma(E)$, from which we extract the carrier mobility,
\begin{equation}
\mu = \frac{1}{e} \frac{\x{d}\sigma}{\x{d}n},
\end{equation}
where $n$ is the carrier density obtained by integrating the density of states. In our simulations, we employ a standard nearest-neighbour tight-binding Hamiltonian for graphene, with the puddles modelled as a set of Gaussian electrostatic potentials. To reproduce prior STS measurements of graphene on SiO$_2$, we set the width of each puddle to $10$ nm, the height of each puddle is randomly chosen within $[-W,W]$ with $W = 50$ meV, and the puddle density is $0.04\%$.
\\

The puddle height scales inversely with the average dielectric environment, $W = W_0/\epsilon_\x{avg}$, where $\epsilon_\x{avg} = (\epsilon_\x{top} + \epsilon_\x{bot})/2$
is the average permittivity of the media above and below the graphene layer \cite{Adam2011, Sharma2015}. Here, the value of $W_0$ is chosen to correspond to STS measurements of graphene on SiO$_2$ \cite{Xue2011}. 
In our simulations, we consider puddle heights of $W = 50$, $25$, and $10$ meV, corresponding to $\epsilon_\x{top} = 1$, $5.9$, and $20.6$. We also consider larger puddle heights of $W = 100$, $200$, and $400$ meV. These latter values correspond to $\epsilon_\x{top} = 1$, but with the graphene much dirtier.

\begin{figure}[tbh]
    \centering
	 \includegraphics[width=\linewidth]{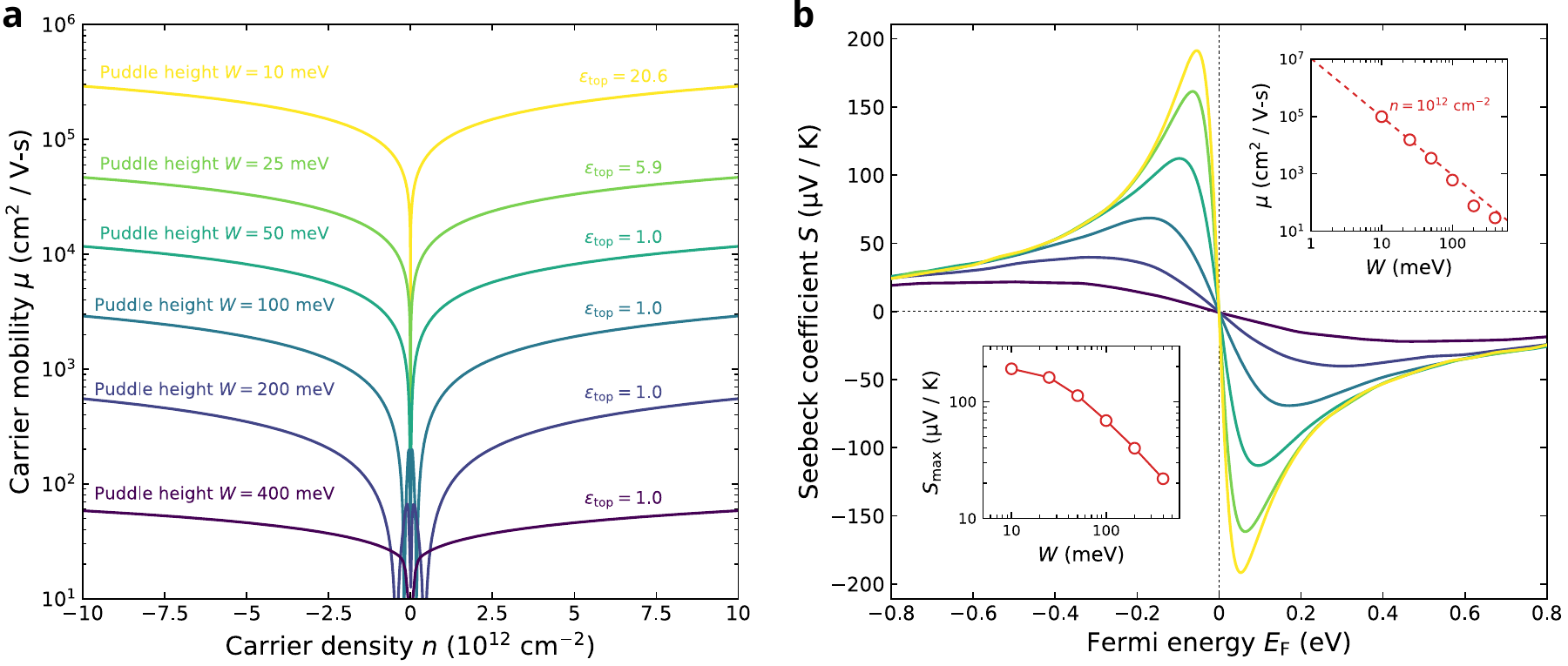}
	\caption{\textbf{Tuning electronic transport in graphene by modulating the electron-hole puddle height via dielectric screening.} \textbf{a,} Carrier mobility as a function of carrier density. \textbf{b,} Seebeck coefficient as a function of Fermi energy, for different puddle heights. Upper right inset: scaling of the mobility with puddle height at $n = 10^{12}$ cm$^{-2}$. The dashed line indicates the scaling $\mu \propto 1/W^2$. Lower left inset: scaling of the maximal value of $S$ with puddle height.}
	\label{fig:theory_transport}
\end{figure}

Figure \ref{fig:theory_transport}a shows the density-dependent mobility for each puddle height. Reducing the puddle height has a clear impact on charge transport; reducing $W$ from 50 to 10 meV ($\epsilon_\x{top} = 1 \to 20.6$) increases the mobility by a factor of approximately $25$ at high densities, from $10\,000$ to more than $200\,000$ cm$^2$/V-s.
The upper right inset of Fig.\ \ref{fig:theory_transport}b shows how the mobility scales with puddle height at a carrier density of $n = 10^{12}$ cm$^{-2}$. The open symbols are numerical data, and the dashed line indicates quadratic scaling, $\mu \propto 1/W^2$. In a recent experimental work, Domaretskiy and coworkers demonstrated that close proximity ($\approx 1$ nm) to a graphite gate could reduce electrostatic potential fluctuations in graphene down to below $1$ meV, allowing them to measure mobilities reaching above $10^7$ cm$^2$/V-s \cite{Domaretskiy2025}. Following the scaling trend we see in the inset, reducing the puddle height down to less than $1$ meV also yields $\mu > 10^7$ cm$^2$/V-s, consistent with their measurements.
\\

Next we examine how the Seebeck coefficient of graphene is altered by dielectric screening. This is calculated from the conductivity as \cite{Sivan1986}
\begin{gather}
	S(E) = -\frac{1}{eT} \frac{K_1(E)}{K_0(E)}, \label{eq_seebeck}
\end{gather}
where $K_j(E) = \int \left( E - \epsilon \right)^j \sigma(\epsilon) \left( -\partial f / \partial \epsilon \right) d\epsilon$, $T = 300$ K is the ambient temperature, and $f$ is the Fermi-Dirac distribution. The main panel of Fig.\ \ref{fig:theory_transport}b shows the Seebeck coefficient of graphene for each puddle height. As expected, $S$ increases with decreasing puddle height, suggesting that dielectric engineering may be an efficient means of tailoring graphene devices using this effect, such as photodetectors based on the photo-thermoelectric effect \cite{Lemme2011, Song2011, Cai2014, Schuler2016, Muench2019, Cummings2021, Koepfli2024, Vicarelli2012b, Castilla2019, Viti2020, Soundarapandian2026}.
\\

The lower left inset shows how the maximum value of $S$ scales with puddle height. Increasing the top dielectric from $\epsilon_\x{top} = 1 \to 20.6$ enhances $S$ by $70 \%$, from $113$ to $192$ $\upmu$V/K. Another interesting feature is that $S$ appears to saturate with decreasing puddle height, approaching approximately $200$ $\upmu$V/K. This value is similar to measurements of graphene on hBN ($S \approx 180$ $\upmu$V/K) \cite{Duan2016}, suggesting an upper achievable value of $S$ in uniform single-layer graphene at room temperature, i.e., in the absence of engineered junctions \cite{Hwang2023} or hydrodynamic effects at lower temperatures \cite{Foster2009}.
\\

\section*{Discussion}

Using pump-probe measurements of graphene in different liquid environments, we observed systematically slower rise and decay dynamics of photoexcited carriers as the dielectric constant of the liquid increased. We ascribe these observations to the screening effect of the liquids on carrier-carrier interactions in graphene. This screening leads to less efficient carrier heating, as a larger fraction of initially excited carriers relax through phonon emission rather than through carrier-carrier interactions. This means that less energy stays within the electron system. Since carrier-carrier interactions are also crucial for re-thermalization during cooling, increased screening also slows cooling. Moreover, increased screening results in a longer persistence of net photoexcited charge carriers in graphene. In addition to controlling carrier dynamics, screening also affects electronic and thermoelectric properties, as we showed through linear-scaling quantum transport simulations. In particular, increased screening leads to charge mobilities above 10$^7$ cm$^2$/V-s and Seebeck coefficients up to 200 $\upmu$V/K.
\\

These results are of fundamental interest because they offer an external mechanism for control of intrinsic graphene dynamics and transport, which govern its optoelectronic, electronic, and thermoelectric properties. The ability to control these properties is also promising for applications such as photo-thermoelectric photodetectors. Considering a simple photodetector with a rectangular geometry containing a graphene channel and two contacts on either side of the channel, where a \textit{pn}-junction is created, the photocurrent is given by \cite{Tielrooij2018,Castilla2019,Massicotte2021}

\begin{gather}
    I_{\rm PTE} = \frac{(S_1 - S_2) \Delta T}{R} , \\
    \Delta T = \frac{P_{\rm abs} \tau_{\rm cool}}{A_{\rm active} C_{\rm el}} .
\end{gather} 

Here, $S_1$ and $S_2$ are the Seebeck coefficients in the $p$- and $n$-doped regions, $R$ is the electrical resistance across the channel, $P_{\rm abs}$ is the absorbed optical power, $\tau_{\rm cool}$ is the carrier cooling time, $A_{\rm active}$ is the active area where the carrier temperature is higher than the ambient temperature, and $C_{\rm el}$ is the electronic heat capacity. By embedding graphene in an environment with a higher dielectric constant, the Seebeck coefficients will increase, leading to a larger factor $(S_1 - S_2)$. The charge mobility is also increased, leading to a reduced device resistance $R$. The carrier cooling time $\tau_{\rm cool}$ increases due to slower cooling caused by screened carrier-carrier interactions. Furthermore, the reduced puddle energy allows tuning the Fermi energy very close to the Dirac point, where the electronic heat capacity is very small, thereby increasing the electron temperature. These factors all contribute to an increased photoresponse. However, there is a trade-off: the heating efficiency can be reduced as a result of screened electron-electron interactions, meaning that some initial energy is lost to phonons before the electronic system thermalizes. Moreover, slower cooling also leads to a longer cooling length and therefore a larger active area $A_{\rm active}$, which is not necessarily advantageous for photo-thermoelectric photodetectors. Since $A_{\rm active} \propto \sqrt{\tau_{\rm cool}}$, the net effect is still an increased photoresponse. 
\\

Future work can explore these trade-offs for photodetectors and other applications, such as nonlinear photonic devices based on graphene. It will also be important to explore alternative approaches to tune the dielectric environment, as liquids are not easily compatible with current device processing. Nearby metallic gates or high-permittivity dielectric materials could be good options, provided they do not host phonon modes that act as efficient heat sinks for hot electrons in graphene. Another interesting observation is that photoexcited carriers remain in the conduction band longer than in the screened case. This ``semiconductor-like'' behaviour opens avenues to exploiting alternative photoresponse mechanisms and functionalities that were not possible with the usual ``metal-like'' behaviour of graphene. 
\\

\section*{Methods}
\label{meth}

\textbf{Experimental details} 

To perform optical pump -- THz probe measurements on graphene with different liquid environments, we adapted a silica flow cell, such that one of the windows can be removed in order transfer graphene onto it. For this transfer, we used wet transfer of CVD graphene on copper foil, transferred using cellulose acetate butyrate (CAB) dissolved in ethyl acetate, which we spin-coated on top of graphene. We first etched the copper in an aqueous solution of 3g/100 ml ammonium persulfate, which was filtered by a 0.2-$\upmu$m Nylon membrane filter. After the transferred CAB/graphene on SiO$_2$ had completely dried, we removed the CAB in an acetone bath for 12 hours. We carried out all steps in dust-free environments. The flow cell was prepared in such a way that the thickness of the solvent superstrate was 50 $\upmu$m, as defined by the thickness of two sides of the window. This thickness ensured sufficient THz transmission. During the studies of the effect of the dielectric environment, we injected a solvent of interest into the silica flow cell using a syringe. To exclude the effect of inhomogeneity of graphene and ensure a reliable comparison, we conducted all the THz measurements at a single spot of the graphene sample, while varying only the solvent that was injected into the silica flow cell. While the dielectric environment above the graphene varied depending on the injected gas or solvent, the dielectric constant below the graphene was fixed to $\epsilon$ = 3.9, due to the fixed quartz substrate. 
\\

\noindent \textbf{Transport simulations}

To simulate charge transport in graphene with disorder on the length scale of electron-hole puddles, we employ a linear-scaling quantum transport method capable of handling systems with many millions of atoms \cite{Fan2021}. With this method, we first calculate the time- and energy-dependent mean-square displacement (MSD) of an initial state $\ket{\psi}$,
\begin{gather}
	\x{MSD}(E,t) = \Delta X^2(E,t) + \Delta Y^2(E,t), \\
	\Delta X^2 / \Delta Y^2 (E,t) = \frac{ \braket{\psi_{X/Y}(t) | \delta(E-\hat{H}) | \psi_{X/Y}(t)} }{ \braket{\psi | \delta(E-\hat{H}) | \psi} },
\end{gather}
where $\ket{\psi(t)} = \hat{U}(t) \ket{\psi}$, $\ket{\psi_{X/Y}(t)} = [\hat{X}/\hat{Y},\hat{U}(t)] \ket{\psi}$, $\hat{X}$ ($\hat{Y}$) is the position operator along the $x$ ($y$) direction, $\hat{U}(t) = \exp(-\x{i} \hat{H} t / \hbar)$ is the time evolution operator, and $\hat{H}$ is the tight-binding Hamiltonian of graphene,
\begin{gather}
	\hat{H} = t \sum_{\langle i,j \rangle} \hat{c}_{i}^\dagger \hat{c}_{j}
	+ \sum_{i} \left[ \sum_{j=1}^{N_\x{eh}} V_j \exp\left(-\frac{|\bm{r}_i-\bm{r}_j|^2}{2\xi^2}\right) \right] \hat{c}_{i}^\dagger \hat{c}_{i}.
\end{gather}

The first term in $\hat{H}$ is the nearest-neighbor hopping with $t = -2.7$ eV. The second term captures the effect of electron-hole puddles, with each puddle modeled as a Gaussian variation of the onsite electrostatic potential \cite{Adam2009, Klos2010}. There are a total of $N_\x{eh}$ puddles, each with random center $\bm{r}_j$, random height $V_j \in [-W,W]$, and uniform spatial width $\xi = 10$ nm. Here we let $N_\x{eh}/N = 0.04\%$, with $N = 8 \times 10^6$ the number of atoms in our simulated system.

Rather than diagonalizing the Hamiltonian, the operators $\hat{U}(t)$ and $\delta(E-\hat{H})$ are expanded as a series of Chebyshev polynomials using the kernel polynomial method \cite{Weisse2006}. Here we use the Jackson kernel and $7000$ polynomials, corresponding to an energy broadening of $6$ meV. The initial state is chosen as a random-phase state in the site basis $\ket{\psi} = \frac{1}{\sqrt{N}} [\x{e}^{\x{i}\xi_1} \, ... \, \x{e}^{\x{i}\xi_N}]^T$, where $\xi_n$ is a random number evenly distributed in $[0,2\pi)$. To reduce numerical noise, we average over $10$ initial states and puddle configurations.
From the MSD, we then calculate the diffusivity and the conductivity as
\begin{gather}
	D(E,t) = \frac{1}{4} \frac{\partial}{\partial t} \x{MSD}(E,t), \\
	\sigma(E) = e^2 \rho(E) D_\x{sat}(E),
\end{gather}
where $\rho$ is the density of states and $D_\x{sat}$ is the saturated value of $D$ at long times.

\section*{Acknowledgments}

X.J. acknowledges the financial support by DFG through the Excellence Initiative by the Graduate School of Excellence Materials Science in Mainz (MAINZ) (GSC 266) and support from the Max Planck Graduate Center mit der Johannes Gutenberg-Universität Mainz (MPGC). The ICN2 is funded by the CERCA programme / Generalitat de Catalunya. The ICN2 is supported by the Severo Ochoa Centres of Excellence programme, Grant CEX2021-001214-S, funded by MCIU/AEI/10.13039.501100011033. K.-J.T. acknowledges funding from the European Union’s Horizon 2020 research and innovation program under Grant Agreement No. 101125457 (ERC CoG ``EQUATE''), Spanish MCIN/AEI project PID2022-142730NB-I00 ``HYDROPTO'', and Flag-ERA grant ENPHOCAL, by MICIN with No. PCI2021-122101-2A
(Spain).

\clearpage
\bibliography{references}

@article{Sivan1986,
author =
{
	Sivan, U.
	and Imry, Y.
},
title = {{Multichannel Landauer formula for thermoelectric transport with application to thermopower near the mobility edge}},
journal = {Phys. Rev. B},
volume = {33},
pages = {551--558},
year = {1986},
doi = {10.1103/PhysRevB.33.551},
}

@article{Lemme2011,
author =
{
	Lemme, Max C.
	and Koppens, Frank H. L.
	and Falk, Abram L.
	and Rudner, Mark S.
	and Park, Hongkun
	and Levitov, Leonid S.
	and Marcus, Charles M.
},
title = {{Gate-Activated Photoresponse in a Graphene p–n Junction}},
journal = {Nano Lett.},
volume = {11},
pages = {4134--4137},
year = {2011},
doi = {10.1021/nl2019068},
}

@article{Song2011,
author =
{
	Song, Justin C. W.
	and Rudner, Mark S.
	and Marcus, Charles M.
	and Levitov, Leonid S.
},
title = {{Hot Carrier Transport and Photocurrent Response in Graphene}},
journal = {Nano Lett.},
volume = {11},
pages = {4688--4692},
year = {2011},
doi = {10.1021/nl202318u},
}

@article{Cai2014,
author =
{
	Cai, Xinghan
	and Sushkov, Andrei B.
	and Suess, Ryan J.
	and Jadidi, Mohammad M.
	and Jenkins, Gregory S.
	and Nyakiti, Luke O.
	and Myers-Ward, Rachael L.
	and Li, Shanshan
	and Yan, Jun
	and Gaskill, D. Kurt
	and Murphy, Thomas E.
	and Drew, H. Dennis
	and Fuhrer, Michael S.
},
title = {{Sensitive room-temperature terahertz detection via the photothermoelectric effect in graphene}},
journal = {Nat. Nanotechnol.},
volume = {9},
pages = {814--819},
year = {2014},
doi = {10.1038/nnano.2014.182},
}

@article{Schuler2016,
author =
{
	Schuler, S.
	and Schall, D.
	and Neumaier, D.
	and Dobusch, L.
	and Bethge, O.
	and Schwarz, B.
	and Krall, M.
	and Mueller, T.
},
title = {{Controlled Generation of a p-n Junction in a Waveguide Integrated Graphene Photodetector}},
journal = {Nano Lett.},
volume = {16},
pages = {7107--7112},
year = {2016},
doi = {10.1021/acs.nanolett.6b03374},
}

@article{Muench2019,
author =
{
    Muench, Jakob E.
	and Ruocco, Alfonso
	and Giambra, Marco A.
	and Miseikis, Vaidotas
	and Zhang, Dengke
	and Wang, Junjia
	and Watson, Hannah F. Y.
	and Park, Gyeong C.
	and Akhavan, Shahab
	and Sorianello, Vito
	and Midrio, Michele
	and Tomadin, Andrea
	and Coletti, Camilla
	and Romagnoli, Marco
	and Ferrari, Andrea C.
	and Goykhman, Ilya
},
title = {{Waveguide-Integrated, Plasmonic Enhanced Graphene Photodetectors}},
journal = {Nano Lett.},
volume = {19},
pages = {7632--7644},
year = {2019},
doi = {10.1021/acs.nanolett.9b02238},
}

@article{Cummings2021,
author =
{
    Antidormi, Aleandro
    and Cummings, Aron W.
},
title = {{Optimizing the Photothermoelectric Effect in Graphene}},
journal = {Phys. Rev. Appl.},
volume = {15},
pages = {054049},
year = {2021},
doi = {10.1103/PhysRevApplied.15.054049},
}

@article{Adam2009,
author =
{
    Adam, Shaffique
    and Brouwer, Piet W.
    and Das Sarma, S.
},
title = {{Crossover from quantum to Boltzmann transport in graphene}},
journal = {Phys. Rev. B},
volume = {79},
pages = {201404},
year = {2009},
doi = {10.1103/PhysRevB.79.201404},
}

@article{Klos2010,
author =
{
    K\l{}os, J. W.
    and Zozoulenko, I. V.
},
title = {{Effect of short- and long-range scattering on the conductivity of graphene: Boltzmann approach vs tight-binding calculations}},
journal = {Phys. Rev. B},
volume = {82},
pages = {081414},
year = {2010},
doi = {10.1103/PhysRevB.82.081414},
}

@article{Deshpande2009,
author =
{
    Deshpande, A.
    and Bao, W.
    and Miao, F.
    and Lau, C. N.
    and LeRoy, B. J.
},
title = {{Spatially resolved spectroscopy of monolayer graphene on SiO$_2$}},
journal = {Phys. Rev. B},
volume = {79},
pages = {205411},
year = {2009},
doi = {10.1103/PhysRevB.79.205411},
}

@Article{Xue2011,
author =
{
    Xue, Jiamin
    and Sanchez-Yamagishi, Javier
    and Bulmash, Danny
    and Jacquod, Philippe
    and Deshpande, Aparna
    and Watanabe, K.
    and Taniguchi, T.
    and Jarillo-Herrero, Pablo
    and LeRoy, Brian J.
},
title = {{Scanning tunnelling microscopy and spectroscopy of ultra-flat graphene on hexagonal boron nitride}},
journal = {Nat. Mater.},
volume = {10},
pages = {282-285},
doi = {10.1038/nmat2968},
year = {2011},
}

@article{Fan2021,
author =
{
    Zheyong Fan
    and José H. Garcia
    and Aron W. Cummings
    and Jose Eduardo Barrios-Vargas
    and Michel Panhans
    and Ari Harju
    and Frank Ortmann
    and Stephan Roche
},
title = {{Linear scaling quantum transport methodologies}},
journal = {Phys. Rep.},
volume = {903},
pages = {1-69},
year = {2021},
doi = {https://doi.org/10.1016/j.physrep.2020.12.001},
}

@article{Weisse2006,
author =
{
    Wei\ss{}e, Alexander
    and Wellein, Gerhard
    and Alvermann, Andreas
    and Fehske, Holger
},
title = {{The kernel polynomial method}},
journal = {Rev. Mod. Phys.},
volume = {78},
pages = {275--306},
year = {2006},
doi = {10.1103/RevModPhys.78.275},
}

@article{Adam2011,
author =
{
    Adam, S.
    and Jung, Suyong
    and Klimov, Nikolai N.
    and Zhitenev, Nikolai B.
    and Stroscio, Joseph A.
    and Stiles, M. D.
},
title = {{Mechanism for puddle formation in graphene}},
journal = {Phys. Rev. B},
volume = {84},
pages = {235421},
year = {2011},
doi = {10.1103/PhysRevB.84.235421},
}

@article{Sharma2015,
author =
{
    Sharma, P.
    and Mišković, Z. L.
},
title = {{Ionic screening of charged impurities in electrolytically gated graphene: A partially linearized Poisson-Boltzmann model}},
journal = {J. Chem. Phys.},
volume = {143},
pages = {134118},
year = {2015},
doi = {10.1063/1.4932179},
}

@article{Domaretskiy2025,
author = 
{
    Domaretskiy, Daniil
    and Wu, Zefei
    and Nguyen, Van Huy
    and Hayward, Ned
    and Babich, Ian
    and Li, Xiao
    and Nguyen, Ekaterina
    and Barrier, Julien
    and Indykiewicz, Kornelia
    and Wang, Wendong
    and Gorbachev, Roman V.
    and Xin, Na
    and Watanabe, Kenji
    and Taniguchi, Takashi
    and Hague, Lee
    and Fal'ko, Vladimir I.
    and Grigorieva, Irina V.
    and Ponomarenko, Leonid A.
    and Berdyugin, Alexey I.
    and Geim, Andre K.
},
title = {{Proximity screening greatly enhances electronic quality of graphene}},
journal = {Nature},
volume = {644},
pages={646--651},
year = {2025},
doi = {10.1038/s41586-025-09386-0},
}

@article{Duan2016,
author =
{
    Junxi Duan
    and Xiaoming Wang
    and Xinyuan Lai
    and Guohong Li
    and Kenji Watanabe
    and Takashi Taniguchi
    and Mona Zebarjadi
    and Eva Y. Andrei
},
title = {{High thermoelectric power factor in graphene/hBN devices}},
journal = {Proc. Natl. Acad. Sci. U.S.A.},
volume = {113},
pages = {14272-14276},
year = {2016},
doi = {10.1073/pnas.1615913113},
}

@article{Hwang2023,
author =
{
    Hyeon Jun Hwang
    and So-Young Kim
    and Sang Kyung Lee
    and Byoung Hun Lee
},
title = {{Large scale graphene thermoelectric device with high power factor using gradient doping profile}},
journal = {Carbon},
volume = {201},
pages = {467-472},
year = {2023},
doi = {https://doi.org/10.1016/j.carbon.2022.09.048},
}

@article{Foster2009,
author = {Foster, Matthew S. and Aleiner, Igor L.},
title = {{Slow imbalance relaxation and thermoelectric transport in graphene}},
journal = {Phys. Rev. B},
volume = {79},
pages = {085415},
year = {2009},
doi = {10.1103/PhysRevB.79.085415},
}

@article{Martinez2013,
author = {Martinez, Amos and Sun, Zhipei},
doi = {10.1038/nphoton.2013.304},
issn = {1749-4885},
journal = {Nature Photonics},
month = {nov},
number = {11},
pages = {842--845},
title = {{Nanotube and graphene saturable absorbers for fibre lasers}},
volume = {7},
year = {2013}
}

@article{Autere2018,
author = {Autere, Anton and Jussila, Henri and Dai, Yunyun and Wang, Yadong and Lipsanen, Harri and Sun, Zhipei},
doi = {10.1002/adma.201705963},
issn = {15214095},
journal = {Advanced Materials},
month = {jun},
number = {24},
pages = {1705963},
pmid = {21973122},
title = {{Nonlinear Optics with 2D Layered Materials}},
volume = {30},
year = {2018}
}

@article{Tan2020,
author = {Tan, Teng and Jiang, Xiantao and Wang, Cong and Yao, Baicheng and Zhang, Han},
doi = {10.1002/advs.202000058},
issn = {2198-3844},
journal = {Advanced Science},
month = {jun},
number = {11},
pages = {2000058},
title = {{2D Material Optoelectronics for Information Functional Device Applications: Status and Challenges}},
volume = {7},
year = {2020}
}

@article{Castilla2019,
archivePrefix = {arXiv},
arxivId = {1905.01881},
author = {Castilla, Sebasti{\'{a}}n and Terr{\'{e}}s, Bernat and Autore, Marta and Viti, Leonardo and Li, Jian and Nikitin, Alexey Y. and Vangelidis, Ioannis and Watanabe, Kenji and Taniguchi, Takashi and Lidorikis, Elefterios and Vitiello, Miriam S. and Hillenbrand, Rainer and Tielrooij, Klaas-Jan and Koppens, Frank H.L.},
doi = {10.1021/acs.nanolett.8b04171},
eprint = {1905.01881},
issn = {1530-6984},
journal = {Nano Letters},
month = {may},
number = {5},
pages = {2765--2773},
pmid = {30882226},
title = {{Fast and Sensitive Terahertz Detection Using an Antenna-Integrated Graphene pn Junction}},
volume = {19},
year = {2019}
}

@article{Vicarelli2012b,
archivePrefix = {arXiv},
arxivId = {1203.3232},
author = {Vicarelli, L. and Vitiello, M. S. and Coquillat, D. and Lombardo, A. and Ferrari, A. C. and Knap, W. and Polini, M. and Pellegrini, V. and Tredicucci, A.},
doi = {10.1038/nmat3417},
eprint = {1203.3232},
issn = {1476-1122},
journal = {Nature Materials},
month = {oct},
number = {10},
pages = {865--871},
pmid = {22961203},
publisher = {Nature Publishing Group},
title = {{Graphene field-effect transistors as room-temperature terahertz detectors}},
volume = {11},
year = {2012}
}

@article{Viti2020,
author = {Viti, Leonardo and Purdie, David G. and Lombardo, Antonio and Ferrari, Andrea C. and Vitiello, Miriam S.},
doi = {10.1021/acs.nanolett.9b05207},
issn = {1530-6984},
journal = {Nano Letters},
month = {may},
number = {5},
pages = {3169--3177},
pmid = {32301617},
title = {{HBN-Encapsulated, Graphene-based, Room-temperature Terahertz Receivers, with High Speed and Low Noise}},
volume = {20},
year = {2020}
}

@article{Soundarapandian2026,
archivePrefix = {arXiv},
arxivId = {2411.02269},
author = {Soundarapandian, Karuppasamy Pandian and Castilla, Sebasti{\'{a}}n and Koepfli, Stefan M. and Marconi, Simone and Kulmer, Laurenz and Vangelidis, Ioannis and de la Bastida, Ronny and Rongione, Enzo and Terr{\'{e}}s, Bernat and Tongay, Seth Ariel and Watanabe, Kenji and Taniguchi, Takashi and Lidorikis, Elefterios and Tielrooij, Klaas-Jan and Leuthold, Juerg and Koppens, Frank H. L.},
doi = {10.1038/s41467-026-69186-6},
eprint = {2411.02269},
isbn = {4146702669186},
issn = {2041-1723},
journal = {Nature Communications},
month = {mar},
number = {1},
pages = {2627},
title = {{High-speed graphene-based sub-terahertz receivers enabling wireless communications for 6G and beyond}},
volume = {17},
year = {2026}
}

@article{Koepfli2024,
author = {Koepfli, Stefan M. and Baumann, Michael and Gadola, Robin and Nashashibi, Shadi and Koyaz, Yesim and Rieben, Daniel and G{\"{u}}ng{\"{o}}r, Arif Can and Doderer, Michael and Keller, Killian and Fedoryshyn, Yuriy and Leuthold, Juerg},
doi = {10.1038/s41467-024-51599-w},
issn = {2041-1723},
journal = {Nature Communications},
month = {aug},
number = {1},
pages = {7351},
title = {{Controlling photothermoelectric directional photocurrents in graphene with over 400 GHz bandwidth}},
volume = {15},
year = {2024}
}

@article{Soavi2018a,
archivePrefix = {arXiv},
arxivId = {1710.03694},
author = {Soavi, Giancarlo and Wang, Gang and Rostami, Habib and Purdie, David G and {De Fazio}, Domenico and Ma, Teng and Luo, Birong and Wang, Junjia and Ott, Anna K and Yoon, Duhee and Bourelle, Sean A and Muench, Jakob E and Goykhman, Ilya and {Dal Conte}, Stefano and Celebrano, Michele and Tomadin, Andrea and Polini, Marco and Cerullo, Giulio and Ferrari, Andrea C},
doi = {10.1038/s41565-018-0145-8},
eprint = {1710.03694},
isbn = {4156501801458},
issn = {1748-3387},
journal = {Nature Nanotechnology},
month = {jul},
number = {7},
pages = {583--588},
pmid = {29784965},
publisher = {Springer US},
title = {{Broadband, electrically tunable third-harmonic generation in graphene}},
volume = {13},
year = {2018}
}

@article{Hafez2018,
author = {Hafez, Hassan A. and Kovalev, Sergey and Deinert, Jan-Christoph and Mics, Zolt{\'{a}}n and Green, Bertram and Awari, Nilesh and Chen, Min and Germanskiy, Semyon and Lehnert, Ulf and Teichert, Jochen and Wang, Zhe and Tielrooij, Klaas-Jan and Liu, Zhaoyang and Chen, Zongping and Narita, Akimitsu and M{\"{u}}llen, Klaus and Bonn, Mischa and Gensch, Michael and Turchinovich, Dmitry},
doi = {10.1038/s41586-018-0508-1},
issn = {0028-0836},
journal = {Nature},
month = {sep},
number = {7724},
pages = {507--511},
pmid = {30202091},
publisher = {Springer US},
title = {{Extremely efficient terahertz high-harmonic generation in graphene by hot Dirac fermions}},
volume = {561},
year = {2018}
}

@article{Jiang2018,
author = {Jiang, Tao and Huang, Di and Cheng, Jinluo and Fan, Xiaodong and Zhang, Zhihong and Shan, Yuwei and Yi, Yangfan and Dai, Yunyun and Shi, Lei and Liu, Kaihui and Zeng, Changgan and Zi, Jian and Sipe, J. E. and Shen, Yuen-Ron and Liu, Wei-Tao and Wu, Shiwei},
doi = {10.1038/s41566-018-0175-7},
issn = {1749-4885},
journal = {Nature Photonics},
month = {jul},
number = {7},
pages = {430--436},
title = {{Gate-tunable third-order nonlinear optical response of massless Dirac fermions in graphene}},
volume = {12},
year = {2018}
}

@article{Luo2019,
author = {Luo, Fang and Fan, Yansong and Peng, Gang and Xu, Shuigang and Yang, Yaping and Yuan, Kai and Liu, Jinxin and Ma, Wei and Xu, Wei and Zhu, Zhi Hong and Zhang, Xue-Ao and Mishchenko, Artem and Ye, Yu and Huang, Han and Han, Zheng and Ren, Wencai and Novoselov, Kostya S and Zhu, Mengjian and Qin, Shiqiao},
doi = {10.1021/acsphotonics.9b00667},
issn = {2330-4022},
journal = {ACS Photonics},
month = {aug},
number = {8},
pages = {2117--2125},
title = {{Graphene Thermal Emitter with Enhanced Joule Heating and Localized Light Emission in Air}},
volume = {6},
year = {2019}
}

@article{Shiue2019,
author = {Shiue, Ren-Jye and Gao, Yuanda and Tan, Cheng and Peng, Cheng and Zheng, Jiabao and Efetov, Dmitri K. and Kim, Young Duck and Hone, James and Englund, Dirk},
doi = {10.1038/s41467-018-08047-3},
issn = {2041-1723},
journal = {Nature Communications},
month = {dec},
number = {1},
pages = {109},
pmid = {30631048},
publisher = {Nature Publishing Group},
title = {{Thermal radiation control from hot graphene electrons coupled to a photonic crystal nanocavity}},
volume = {10},
year = {2019}
}

@article{Kim2015,
author = {Kim, Young Duck and Kim, Hakseong and Cho, Yujin and Ryoo, Ji Hoon and Park, Cheol-Hwan and Kim, Pilkwang and Kim, Yong Seung and Lee, Sunwoo and Li, Yilei and Park, Seung-Nam and {Shim Yoo}, Yong and Yoon, Duhee and Dorgan, Vincent E. and Pop, Eric and Heinz, Tony F. and Hone, James and Chun, Seung-Hyun and Cheong, Hyeonsik and Lee, Sang Wook and Bae, Myung-Ho and Park, Yun Daniel},
doi = {10.1038/nnano.2015.118},
issn = {1748-3387},
journal = {Nature Nanotechnology},
month = {aug},
number = {8},
pages = {676--681},
publisher = {Nature Publishing Group},
title = {{Bright visible light emission from graphene}},
volume = {10},
year = {2015}
}

@article{Tielrooij2013,
archivePrefix = {arXiv},
arxivId = {arXiv:1210.1205v2},
author = {Tielrooij, K J and Song, J C W and Jensen, S A and Centeno, A and Pesquera, A and {Zurutuza Elorza}, A and Bonn, M and Levitov, L S and Koppens, F H L},
doi = {10.1038/nphys2564},
eprint = {arXiv:1210.1205v2},
isbn = {1745-2473},
issn = {1745-2473},
journal = {Nature Physics},
month = {apr},
number = {4},
pages = {248--252},
title = {{Photoexcitation cascade and multiple hot-carrier generation in graphene}},
volume = {9},
year = {2013}
}

@article{Gierz2013,
archivePrefix = {arXiv},
arxivId = {1304.1389},
author = {Gierz, Isabella and Petersen, Jesse C. and Mitrano, Matteo and Cacho, Cephise and Turcu, I. C. Edmond and Springate, Emma and St{\"{o}}hr, Alexander and K{\"{o}}hler, Axel and Starke, Ulrich and Cavalleri, Andrea},
doi = {10.1038/nmat3757},
eprint = {1304.1389},
issn = {1476-1122},
journal = {Nature Materials},
month = {dec},
number = {12},
pages = {1119--1124},
publisher = {Nature Publishing Group},
title = {{Snapshots of non-equilibrium Dirac carrier distributions in graphene}},
volume = {12},
year = {2013}
}

@article{Brida2013,
archivePrefix = {arXiv},
arxivId = {1209.5729},
author = {Brida, D. and Tomadin, A. and Manzoni, C. and Kim, Y. J. and Lombardo, A. and Milana, S. and Nair, R. R. and Novoselov, K. S. and Ferrari, A. C. and Cerullo, G. and Polini, M.},
doi = {10.1038/ncomms2987},
eprint = {1209.5729},
issn = {2041-1723},
journal = {Nature Communications},
month = {oct},
number = {1},
pages = {1987},
title = {{Ultrafast collinear scattering and carrier multiplication in graphene}},
volume = {4},
year = {2013}
}

@article{Song2013,
archivePrefix = {arXiv},
arxivId = {1209.4346},
author = {Song, Justin C.W. and Tielrooij, Klaas J. and Koppens, Frank H.L. and Levitov, Leonid S.},
doi = {10.1103/PhysRevB.87.155429},
eprint = {1209.4346},
issn = {10980121},
journal = {Physical Review B - Condensed Matter and Materials Physics},
number = {15},
pages = {1--6},
title = {{Photoexcited carrier dynamics and impact-excitation cascade in graphene}},
volume = {87},
year = {2013}
}

@article{Jensen2014,
author = {Jensen, S A and Mics, Z and Ivanov, I and Varol, H S and Turchinovich, D and Koppens, F H L and Bonn, M and Tielrooij, K J},
doi = {10.1021/nl502740g},
issn = {1530-6984},
journal = {Nano Letters},
month = {oct},
number = {10},
pages = {5839--5845},
title = {{Competing Ultrafast Energy Relaxation Pathways in Photoexcited Graphene}},
volume = {14},
year = {2014}
}

@article{Rana2009,
author = {Rana, Farhan and George, Paul A and Strait, Jared H and Dawlaty, Jahan and Shivaraman, Shriram and Chandrashekhar, Mvs and Spencer, Michael G},
doi = {10.1103/PhysRevB.79.115447},
issn = {1098-0121},
journal = {Physical Review B},
month = {mar},
number = {11},
pages = {115447},
title = {{Carrier recombination and generation rates for intravalley and intervalley phonon scattering in graphene}},
volume = {79},
year = {2009}
}

@article{Breusing2011,
author = {Breusing, M. and Kuehn, S. and Winzer, T. and Mali{\'{c}}, E. and Milde, F. and Severin, N. and Rabe, J. P. and Ropers, C. and Knorr, A. and Elsaesser, T.},
doi = {10.1103/PhysRevB.83.153410},
issn = {1098-0121},
journal = {Physical Review B},
month = {apr},
number = {15},
pages = {153410},
title = {{Ultrafast nonequilibrium carrier dynamics in a single graphene layer}},
volume = {83},
year = {2011}
}

@article{Malard2013,
author = {Malard, Leandro M and {Fai Mak}, Kin and {Castro Neto}, A H and Peres, N M R and Heinz, Tony F},
doi = {10.1088/1367-2630/15/1/015009},
issn = {1367-2630},
journal = {New Journal of Physics},
month = {jan},
number = {1},
pages = {015009},
title = {{Observation of intra- and inter-band transitions in the transient optical response of graphene}},
volume = {15},
year = {2013}
}

@article{Song2012,
author = {Song, Justin C W and Reizer, Michael Y and Levitov, Leonid S},
doi = {10.1103/PhysRevLett.109.106602},
issn = {0031-9007},
journal = {Physical Review Letters},
month = {sep},
number = {10},
pages = {106602},
title = {{Disorder-Assisted Electron-Phonon Scattering and Cooling Pathways in Graphene}},
volume = {109},
year = {2012}
}

@article{Graham2013,
author = {Graham, Matt W and Shi, Su-Fei and Wang, Zenghui and Ralph, Daniel C and Park, Jiwoong and McEuen, Paul L},
doi = {10.1021/nl4030787},
issn = {1530-6984},
journal = {Nano Letters},
month = {nov},
number = {11},
pages = {5497--5502},
pmid = {24124889},
title = {{Transient Absorption and Photocurrent Microscopy Show That Hot Electron Supercollisions Describe the Rate-Limiting Relaxation Step in Graphene}},
volume = {13},
year = {2013}
}

@article{Wang2010,
archivePrefix = {arXiv},
arxivId = {0909.4912},
author = {Wang, Haining and Strait, Jared H and George, Paul A and Shivaraman, Shriram and Shields, Virgil B and Chandrashekhar, Mvs and Hwang, Jeonghyun and Rana, Farhan and Spencer, Michael G and Ruiz-Vargas, Carlos S and Park, Jiwoong},
doi = {10.1063/1.3291615},
eprint = {0909.4912},
issn = {0003-6951},
journal = {Applied Physics Letters},
month = {feb},
number = {8},
pages = {081917},
title = {{Ultrafast relaxation dynamics of hot optical phonons in graphene}},
volume = {96},
year = {2010}
}

@article{Tielrooij2018,
archivePrefix = {arXiv},
arxivId = {1702.03766},
author = {Tielrooij, Klaas-Jan Jan and Hesp, Niels C.H. H. and Principi, Alessandro and Lundeberg, Mark B. and Pogna, Eva A.A. A. and Banszerus, Luca and Mics, Zolt{\'{a}}n and Massicotte, Mathieu and Schmidt, Peter and Davydovskaya, Diana and Purdie, David G. and Goykhman, Ilya and Soavi, Giancarlo and Lombardo, Antonio and Watanabe, Kenji and Taniguchi, Takashi and Bonn, Mischa and Turchinovich, Dmitry and Stampfer, Christoph and Ferrari, Andrea C. and Cerullo, Giulio and Polini, Marco and Koppens, Frank H.L. L.},
doi = {10.1038/s41565-017-0008-8},
eprint = {1702.03766},
isbn = {1748-3395 (Electronic) 1748-3387 (Linking)},
issn = {17483395},
journal = {Nature Nanotechnology},
month = {jan},
number = {1},
pages = {41--46},
pmid = {29180742},
title = {{Out-of-plane heat transfer in van der Waals stacks through electron-hyperbolic phonon coupling}},
volume = {13},
year = {2018}
}

@article{Yang2018,
archivePrefix = {arXiv},
arxivId = {1702.02829},
author = {Yang, Wei and Berthou, Simon and Lu, Xiaobo and Wilmart, Quentin and Denis, Anne and Rosticher, Michael and Taniguchi, Takashi and Watanabe, Kenji and F{\`{e}}ve, Gwendal and Berroir, Jean-Marc and Zhang, Guangyu and Voisin, Christophe and Baudin, Emmanuel and Pla{\c{c}}ais, Bernard},
doi = {10.1038/s41565-017-0007-9},
eprint = {1702.02829},
issn = {1748-3387},
journal = {Nature Nanotechnology},
month = {jan},
number = {1},
pages = {47--52},
pmid = {29180743},
title = {{A graphene Zener–Klein transistor cooled by a hyperbolic substrate}},
volume = {13},
year = {2018}
}

@article{Yu2023,
author = {Yu, Xiaoqing and Principi, Alessandro and Tielrooij, Klaas-jan and Bonn, Mischa and Kavokine, Nikita},
doi = {10.1038/s41565-023-01421-3},
isbn = {4156502301421},
issn = {1748-3387},
journal = {Nature Nanotechnology},
month = {aug},
number = {8},
pages = {898--904},
publisher = {Springer US},
title = {{Electron cooling in graphene enhanced by plasmon–hydron resonance}},
volume = {18},
year = {2023}
}

@article{Pogna2021,
archivePrefix = {arXiv},
arxivId = {2103.03527},
author = {Pogna, Eva A. A. and Jia, Xiaoyu and Principi, Alessandro and Block, Alexander and Banszerus, Luca and Zhang, Jincan and Liu, Xiaoting and Sohier, Thibault and Forti, Stiven and Soundarapandian, Karuppasamy and Terr{\'{e}}s, Bernat and Mehew, Jake D. and Trovatello, Chiara and Coletti, Camilla and Koppens, Frank H. L. and Bonn, Mischa and van Hulst, Niek and Verstraete, Matthieu J. and Peng, Hailin and Liu, Zhongfan and Stampfer, Christoph and Cerullo, Giulio and Tielrooij, Klaas-Jan},
eprint = {2103.03527},
journal = {Arxiv},
month = {mar},
title = {{Hot-Carrier Cooling in High-Quality Graphene is Intrinsically Limited by Optical Phonons}},
volume = {2103.03527},
year = {2021}
}

@article{Graham2013a,
author = {Graham, Matt W and Shi, Su-Fei and Wang, Zenghui and Ralph, Daniel C and Park, Jiwoong and McEuen, Paul L},
doi = {10.1021/nl4030787},
issn = {1530-6984},
journal = {Nano Letters},
month = {nov},
number = {11},
pages = {5497--5502},
pmid = {24124889},
title = {{Transient Absorption and Photocurrent Microscopy Show That Hot Electron Supercollisions Describe the Rate-Limiting Relaxation Step in Graphene}},
volume = {13},
year = {2013}
}

@article{Pogna2022,
author = {Pogna, Eva A. A. and Tomadin, Andrea and Balci, Osman and Soavi, Giancarlo and Paradisanos, Ioannis and Guizzardi, Michele and Pedrinazzi, Paolo and Mignuzzi, Sandro and Tielrooij, Klaas-Jan and Polini, Marco and Ferrari, Andrea C. and Cerullo, Giulio},
doi = {10.1021/acsnano.1c04937},
issn = {1936-0851},
journal = {ACS Nano},
month = {mar},
number = {3},
pages = {3613--3624},
title = {{Electrically Tunable Nonequilibrium Optical Response of Graphene}},
volume = {16},
year = {2022}
}

@article{Crossno2016,
archivePrefix = {arXiv},
arxivId = {1509.04713},
author = {Crossno, Jesse and Shi, Jing K. and Wang, K. and Liu, Xiaomeng and Harzheim, Achim and Lucas, Andrew and Sachdev, Subir and Kim, Philip and Taniguchi, Takashi and Watanabe, Kenji and Ohki, Thomas A. and Fong, Kin Chung},
doi = {10.1126/science.aad0343},
eprint = {1509.04713},
issn = {0036-8075},
journal = {Science},
month = {mar},
number = {6277},
pages = {1058--1061},
pmid = {26912362},
title = {{Observation of the Dirac fluid and the breakdown of the Wiedemann-Franz law in graphene}},
volume = {351},
year = {2016}
}

@article{Block2021,
author = {Block, Alexander and Principi, Alessandro and Hesp, Niels C.H. and Cummings, Aron W. and Liebel, Matz and Watanabe, Kenji and Taniguchi, Takashi and Roche, Stephan and Koppens, Frank H.L. and van Hulst, Niek F. and Tielrooij, Klaas Jan},
doi = {10.1038/s41565-021-00957-6},
issn = {17483395},
journal = {Nature Nanotechnology},
number = {11},
pages = {1195--1200},
pmid = {34426681},
publisher = {Springer US},
title = {{Observation of giant and tunable thermal diffusivity of a Dirac fluid at room temperature}},
volume = {16},
year = {2021}
}

@article{Kim2020,
author = {Kim, M and Xu, S G and Berdyugin, A I and Principi, A and Slizovskiy, S and Xin, N and Kumaravadivel, P and Kuang, W and Hamer, M and {Krishna Kumar}, R. and Gorbachev, R V and Watanabe, K and Taniguchi, T and Grigorieva, I V and Fal'ko, V. I. and Polini, M and Geim, A K},
doi = {10.1038/s41467-020-15829-1},
issn = {2041-1723},
journal = {Nature Communications},
month = {dec},
number = {1},
pages = {2339},
title = {{Control of electron-electron interaction in graphene by proximity screening}},
volume = {11},
year = {2020}
}

@book{Madelung1991,
address = {Berlin/Heidelberg},
doi = {10.1007/b44266},
editor = {Madelung, O.},
isbn = {3-540-54417-8},
publisher = {Springer-Verlag},
series = {Landolt-B{\"{o}}rnstein - Group IV Physical Chemistry},
title = {{Static Dielectric Constants of Pure Liquids and Binary Liquid Mixtures}},
volume = {6},
year = {1991}
}

@article{Tomadin2018,
archivePrefix = {arXiv},
arxivId = {1712.02705},
author = {Tomadin, Andrea and Hornett, Sam M. and Wang, Hai I. and Alexeev, Evgeny M. and Candini, Andrea and Coletti, Camilla and Turchinovich, Dmitry and Kl{\"{a}}ui, Mathias and Bonn, Mischa and Koppens, Frank H.L. L. and Hendry, Euan and Polini, Marco and Tielrooij, Klaas-Jan},
doi = {10.1126/sciadv.aar5313},
eprint = {1712.02705},
issn = {2375-2548},
journal = {Science Advances},
month = {may},
number = {5},
pages = {eaar5313},
pmid = {29756035},
title = {{The ultrafast dynamics and conductivity of photoexcited graphene at different Fermi energies}},
volume = {4},
year = {2018}
}

@article{Wu2016f,
archivePrefix = {arXiv},
arxivId = {1603.04934},
author = {Wu, Sanfeng and Wang, Lei and Lai, You and Shan, Wen-Yu and Aivazian, Grant and Zhang, Xian and Taniguchi, Takashi and Watanabe, Kenji and Xiao, Di and Dean, Cory and Hone, James and Li, Zhiqiang and Xu, Xiaodong},
doi = {10.1126/sciadv.1600002},
eprint = {1603.04934},
issn = {2375-2548},
journal = {Science Advances},
month = {may},
number = {5},
pages = {e1600002},
pmid = {27386538},
publisher = {American Association for the Advancement of Science},
title = {{Multiple hot-carrier collection in photo-excited graphene Moir{\'{e}} superlattices}},
volume = {2},
year = {2016}
}

@article{Massicotte2021,
author = {Massicotte, Mathieu and Soavi, Giancarlo and Principi, Alessandro and Tielrooij, Klaas-Jan},
doi = {10.1039/D0NR09166A},
issn = {2040-3364},
journal = {Nanoscale},
number = {18},
pages = {8376--8411},
publisher = {Royal Society of Chemistry},
title = {{Hot carriers in graphene – fundamentals and applications}},
volume = {13},
year = {2021}
}

@article{Wang2017a,
author = {Wang, Hai I. and Braatz, Marie-Luise and Richter, Nils and Tielrooij, Klaas-Jan and Mics, Zoltan and Lu, Hao and Weber, Nils-Eike and M{\"{u}}llen, Klaus and Turchinovich, Dmitry and Kl{\"{a}}ui, Mathias and Bonn, Mischa},
doi = {10.1021/acs.jpcc.7b00347},
issn = {1932-7447},
journal = {The Journal of Physical Chemistry C},
month = {feb},
number = {7},
pages = {4083--4091},
title = {{Reversible Photochemical Control of Doping Levels in Supported Graphene}},
volume = {121},
year = {2017}
}

@article{Frenzel2014,
archivePrefix = {arXiv},
arxivId = {1403.3669},
author = {Frenzel, A. J. and Lui, C. H. and Shin, Y. C. and Kong, J. and Gedik, N.},
doi = {10.1103/PhysRevLett.113.056602},
eprint = {1403.3669},
issn = {0031-9007},
journal = {Physical Review Letters},
month = {jul},
number = {5},
pages = {056602},
title = {{Semiconducting-to-Metallic Photoconductivity Crossover and Temperature-Dependent Drude Weight in Graphene}},
volume = {113},
year = {2014}
}

\end{document}